\documentclass[twocolumn,twoside,slac_two]{revtex4}
\usepackage{graphicx}
\usepackage{fancyhdr}
\usepackage{amsmath,natbib}

\begin{document}

\title{Quasi-Periodic Oscillations in Relativistic Tori}

%

\author{P. Chris Fragile}

\affiliation{Department of Physics, University of California, Santa Barbara, CA 93106, USA}

\begin{abstract}
Motivated by recent interesting work on {\it p}-mode oscillations in 
axisymmetric hydrodynamic 
black-hole tori by Rezzolla, Zanotti, and collaborators, I explore 
the robustness of these oscillations by means of two and three-dimensional 
relativistic hydrodynamic and MHD simulations.  The primary purpose of this 
investigation is to 
determine how the amplitudes of these oscillations are affected by 
the presence of 
known instabilities of black-hole tori, including the Papaloizou-Pringle 
instability (PPI) and the magneto-rotational instability (MRI).  
Both instabilities drive accretion at rates above those considered 
in Rezzolla's work.  The increased accretion can allow wave energy 
to leak out of the torus into the hole.  Furthermore, with the MRI, 
the presence of turbulence, which is absent in the hydrodynamic 
simulations, can lead to turbulent damping (or excitation) of modes.  
The current numerical results are preliminary, but suggest that the 
PPI and MRI both significantly damp acoustic oscillations in tori.
\end{abstract}

\maketitle

\thispagestyle{fancy}


\section{Introduction \label{sec:intro}}

``Diskoseismology'' \citep{now91} 
is a relatively new subfield of accretion disk physics. 
Often diskoseismic oscillations have been overlooked as 
uninteresting because, to be observable from the outside, 
they must be trapped in a particular region of the disk or have a global 
pattern.  For a Keplerian disk in a Newtonian potential, there are 
no global modes and relatively few trapped modes.  Relativistic gravity 
helps the situation somewhat by providing a richer set of trapped modes.  
Nevertheless, even in general relativity, Keplerian thin disks have only a few 
trapped modes, which mostly occur in very restricted regions of the 
disk.

Non-Keplerian disks, on the other hand, have a much richer set of 
diskoseismic modes available for two fundamental reasons.  First, 
in contrast to Keplerian disks, which in principle extend 
to infinity, non-Keplerian disks can be constructed to have a finite size 
and can thus act as a single resonant cavity.  
Second, unlike Keplerian disks, non-Keplerian disks necessarily have 
radial pressure gradients that can provide an 
additional restoring force (along with the centrifugal force and gravity) 
in support of oscillations.

In this work I focus on inertial-acoustic 
({\it p}-mode) oscillations in finite tori.
Two interesting consequences of such oscillations have previously been 
identified.  
First, they can cause periodic changes in the mass quadrupole moment 
of the disks.  For massive tori of nearly nuclear densities, 
this may produce gravitational radiation detectable by LIGO, especially for 
sources located within our galaxy \citep{zan03}.  
Such massive tori could form from 
the gravitational collapse of a massive rotating star or as an 
intermediate state in a binary neutron star merger.
These oscillations may also be important in explaining 
quasi-periodic oscillations (QPOs), particularly 
the harmonically related high-frequency QPOs (HFQPOs) seen in some black-hole 
candidate low-mass X-ray binaries (LMXBs) \citep{rez03}.

My goal here is to extend previous numerical simulations of oscillating 
tori \citep{zan03,zan05} to include magnetic fields and 
nonaxisymmetric perturbations.  I proceed in \S \ref{sec:equations} 
by reviewing the equations of general relativistic magnetohydrodynamics 
as solved in the numerical code used in this work.  The code itself 
is briefly described in \S \ref{sec:method}.  In \S \ref{sec:results} 
I present the results of this study.  I conclude with some final thoughts 
in \S \ref{sec:summary}.

\section{General Relativistic MHD Equations \label{sec:equations}}

This work uses a form of the general relativistic MHD equations similar 
to \citet{dev03a}.  Nevertheless, it is worth explicitly writing these 
equations out for clarity.  In flux-conserving form, the conservation 
equations for mass, internal energy, and momentum and the induction 
equation for magnetic fields are:
\begin{eqnarray}
 \frac{\partial D}{\partial t} + \frac{\partial (DV^i)}{\partial x^i} &=& 0 ,
      \label{eqn:av_de} \\
 \frac{\partial E}{\partial t} + \frac{\partial (EV^i)}{\partial x^i} &=& 
 - P\frac{\partial W}{\partial t} - P\frac{\partial (WV^i)}{\partial x^i} ,
      \label{eqn:av_en} \\
 \frac{\partial S_j}{\partial t} 
 + \frac{\partial (S_j V^i)}{\partial x^i} &=& 
 \frac{\partial (\sqrt{-g} B_j B^0 /{4\pi})}{\partial t} \label{eqn:av_mom} \\ \nonumber
 & & {}+ \frac{\partial ( \sqrt{-g} B_j B^i/{4\pi})}{\partial x^i} \\ \nonumber 
 & & {}+ \frac{1}{2}\left( \frac{S^\mu S^\nu}{S^0} - \frac{\sqrt{-g}}{4\pi} B^\mu B^\nu \right) \frac{\partial g_{\mu\nu}}{\partial x^j} \\ \nonumber
 & & {}- \sqrt{-g} \frac{\partial (P+\vert \vert B \vert \vert^2/{8\pi})}{\partial x^j} , \\
 \frac{\partial \mathcal{B}^i}{\partial t} + \frac{\partial (\mathcal{B}^i V^j) }{\partial x^j} &=& \mathcal{B}^j \frac{\partial V^i}{\partial x^j} , 
      \label{eqn:av_ind}
\end{eqnarray}
where $g$ is the determinant of the 4-metric,
$W=\sqrt{-g} u^t$ is the relativistic boost factor,
$D=W\rho$ is the generalized fluid density,
$P=(\Gamma-1)E/W$ is the fluid pressure,
$V^i=u^i/u^t$ is the transport velocity,
$S_i = W\rho h u_i$ is the covariant momentum density,
and $E=We=W\rho\epsilon$ is the generalized internal energy density.  
There are two representations of the magnetic field in these equations:  
$B^\mu$ is the magnetic field 4-vector ($\vert \vert B \vert \vert^2=g_{\mu \nu} B^\mu B^\nu$) and 
\begin{equation}
\mathcal{B}^\mu = W(B^\mu - B^0 V^\mu)
\end{equation}
is the divergence-free ($\partial \mathcal{B}^i / \partial x^i = 0$), 
spatial ($\mathcal{B}^0=0$) representation of the field.  The time 
component of the magnetic field $B^0$ is recovered from the orthogonality 
condition $B^\mu u_\mu = 0$.

These equations are evolved in
a Kerr-Schild polar coordinate system
$({t},{r},{\theta},{\phi})$, although all of the simulations in this 
work assume a non-rotating ($a=0$) Schwarzschild black hole.
The computational advantages of the ``horizon-adapted''
Kerr-Schild form of the
Kerr metric are described in \citet{pap98} and \citet{fon98b}.
The primary advantage is that, unlike Boyer-Lindquist coordinates,
there are no singularities in the metric terms at the event
horizon.  This is particularly important for numerical calculations
as it allows one to place the grid boundaries inside the horizon,
thus ensuring that they are causally disconnected from the
rest of the flow.

\section{Numerical Method \label{sec:method}}

These simulations are carried out using the numerical code Cosmos++, 
a massively parallel, multidimensional 
(one, two, or three dimensions), adaptive-mesh, 
magnetohydrodynamic code for evolving both 
Newtonian and relativistic flows.  
Cosmos++ employs a time-explicit, operator-split, finite volume 
discretization method with second-order spatial accuracy.  
A detailed description of Cosmos++, including test results, will be 
presented in an upcoming paper.

The simulations are initialized with the analytic solution for
an axisymmetric torus with constant specific angular momentum $l$.
For the initialization, we assume an isentropic equation of state
$P=\kappa \rho^\Gamma$,
although during the evolution, the adiabatic form
$P=(\Gamma-1)E/W$ is used to recover the pressure when solving
equations (\ref{eqn:av_en}) and (\ref{eqn:av_mom}).
We set $\Gamma=4/3$ and $\kappa=0.0229$ 
(in $G=c=1$ units).  

This work presents both two-dimensional (axisymmetric) and 
three-dimensional simulations.  
The two-dimensional simulations are carried out on a grid extending from
$0.98r_{BH}\le r \le r_{max}$ and $0 \le \theta \le \pi$, 
where $r_{BH}=2GM/c^2$ is the radius of the black-hole horizon.  The 
three-dimensional simulations include the full azimuthal range $0 \le
\phi < 2\pi$. We choose $r_{max}$ to be about twice the initial 
outer radius $r_{out}$ of the torus.  In order to
increase the resolution inside the torus, we replace
the radial coordinate ${r}$ with a logarithmic coordinate 
$x_1 = 1 + \ln (r/r_{BH})$ and the angular coordinate $\theta$ with a 
coordinate $x_2$ satisfying $\theta = x_2 + \frac{1}{4} \sin (2x_2)$.  
The grid is resolved with $128\times128$ zones in 2D, 
giving a radial spacing of $0.04 r_G$ near the horizon and $0.6r_G$ near
the outer boundary.  In the present work, 
the 3D simulations are limited to $64^3$ zones.

In the ``background'' regions not determined by the initial
torus solution, we initialize the gas following the spherical Bondi
accretion solution \citep{mic72}.  
We fix the parameters of this solution such that
the rest mass present in the background is negligible
compared to the mass in the torus.  The outer radial boundary
is held fixed with the analytic Bondi solution for all evolved
fields.  The inner radial
boundary uses outflow ($V^r < 0$) boundary conditions.
Data are shared appropriately across angular boundaries.

\section{Results \label{sec:results}}

Results are presented in geometrized units ($G=c=1$) with units of 
length parameterized in terms of the gravitational radius 
of the black hole, $r_G=GM/c^2$.

\subsection{2D Axisymmetric Hydro \label{sec:2dhydro}}

It is instructive to begin this work by reproducing the results 
of a previously published axisymmetric hydrodynamic simulation of an 
oscillating torus.  
This will be useful for illustrating how {\it p}-mode oscillations 
are generally manifested in axisymmetric hydrodynamic tori and 
facilitate easier comparison with 
the subsequent MHD and non-axisymmetric results.  
This is also important as a validation of the new code 
used here.  I choose Model (a) from \citet{zan03} for this purpose.  
In this model, the specific angular momentum, 
which is constant throughout the torus, is $l/M=3.8$.  
The torus just fills the largest closed equipotential surface so that 
the inner torus boundary corresponds with the location of the cusp 
$r_{in}=r_{cusp}=4.576$.  The outer radius for such a torus is 
$r_{out}=15.889$.  
The density center of this torus is located at $r_{center}=8.352$; the 
orbital period at this radius is $t_{orb}=151.7$.  
This will serve as a reference 
dynamical timescale for all these simulations.
To excite the resonant acoustic mode I use an initial radial 
velocity kick.  For convenience this velocity 
is set at a small fraction ($\eta=0.12$) 
of the background Bondi inflow velocity.

\begin{figure}
\scalebox{0.33}{\includegraphics{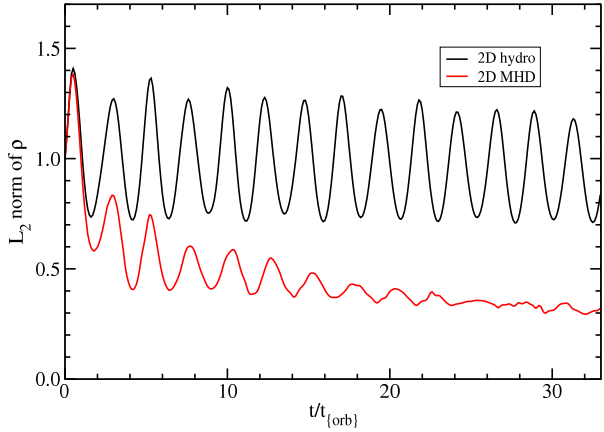}}
\caption{Time history of the $L_2$ norm of the rest-mass density.  
The vertical scale has been adjusted to the initial value of 
$\vert \vert \rho \vert \vert^2$.
\label{fig:l2_rho}}
\end{figure}

\begin{figure}
\scalebox{0.33}{\includegraphics{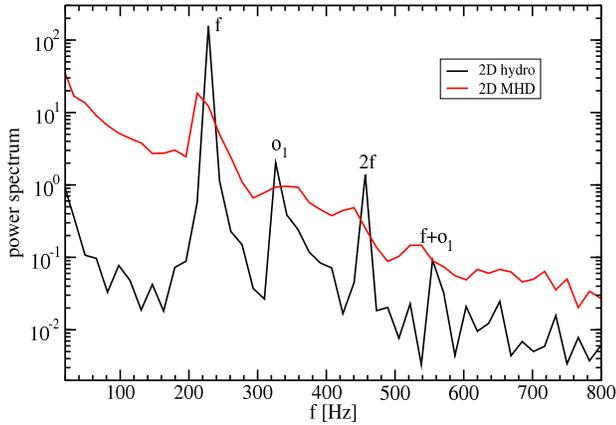}}
\caption{Power spectra of the $L_2$ norm of the rest-mass density.  
The spectra have been arbitrarily normalized such 
that the strongest frequency bin has 
a value of 100.  The frequency is set assuming a black hole mass 
$M_{BH}=2.5 M_\odot$.
\label{fig:powerrho}}
\end{figure}

I use the $L_2$ norm of the rest mass density, defined as 
$\vert\vert \rho \vert \vert^2 = \sum_{i=1}^{N_r} \sum_{j=1}^{N_\theta} \rho_{ij}^2 $, 
to characterized the global oscillatory behavior of the torus.  
The time history and Fourier power spectrum of this quantity are shown 
in Figures \ref{fig:l2_rho} and \ref{fig:powerrho}, respectively.  
The power spectrum, in particular, reveals the rich harmonic structure 
characteristic of diskoseismic modes \citep{kat01}.  
\citet{rez03} also showed that this structure is consistent with 
predictions of linear perturbation analysis of vertically integrated 
relativistic tori.

\begin{figure}
\scalebox{0.33}{\includegraphics{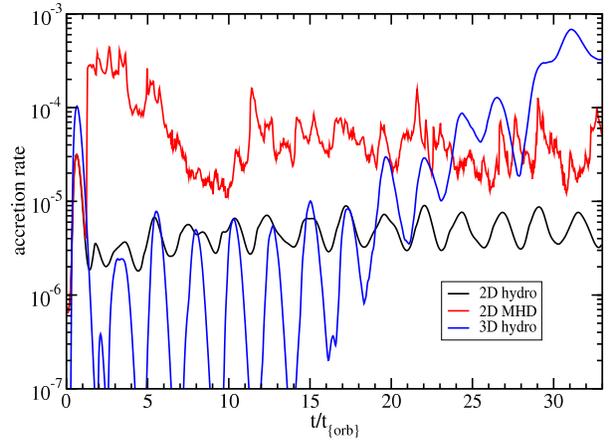}}
\caption{Rest mass accretion rate normalized by the initial mass of 
the torus. 
\label{fig:mdot}}
\end{figure}

\begin{figure}
\scalebox{0.33}{\includegraphics{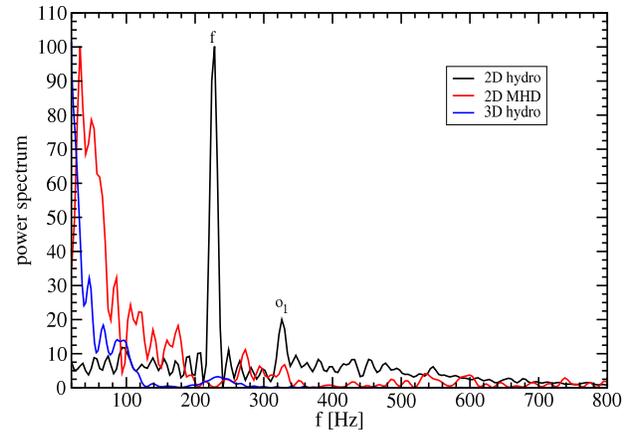}}
\caption{Power spectra of the mass accretion rate.  The spectra have 
been arbitrarily normalized such that the strongest frequency bin has 
a value of 100.  The frequency is set assuming a black hole mass 
$M_{BH}=2.5 M_\odot$.
\label{fig:powermdot}}
\end{figure}

Although the {\it p}-mode of interest in this work is an acoustic wave within 
the disk, the torus parameters for this model are such that this wave 
periodically pushes material over the 
cusp and out of the torus.  This material is accreted into the black hole 
on a dynamical timescale.  Therefore, one manifestation of 
the {\it p}-mode in (marginally stable) tori is a periodic fluctuation 
in the black-hole mass 
accretion rate as shown in Figure \ref{fig:mdot}.  The 
Fourier power spectrum of the accretion history (Figure \ref{fig:powermdot}) 
reveals the same fundamental frequency and first overtone as 
Figure \ref{fig:powerrho}, confirming the connection.

\subsection{3D Hydro \label{sec:3dhydro}}

I now drop the restriction of axisymmetry and consider a 
three-dimensional hydrodynamic torus.  The concern here is that such a 
structure is known to be unstable to low-order non-axisymmetric modes 
(the Papaloizou-Pringle instability or PPI).
Note that this simulation uses a slightly different initial setup than 
the previous one.  First, the velocity perturbation is 
slightly smaller in this case ($\eta=0.06$ instead of 0.12).  In my testing 
I have found that the amplitude of this perturbation has little effect 
on the results, so this difference shouldn't be important.  
This simulation also uses a different set of parameters for the 
background Bondi solution.  These differences account for the much larger 
amplitude of oscillations in the accretion rate for the 
three-dimensional model (Figure \ref{fig:mdot}).  
Nevertheless, the frequency of the oscillations in both 
cases is the same, indicating that the different background treatments do not 
effect the behavior of the {\it p}-mode oscillations.

\begin{figure}
\scalebox{0.33}{\includegraphics{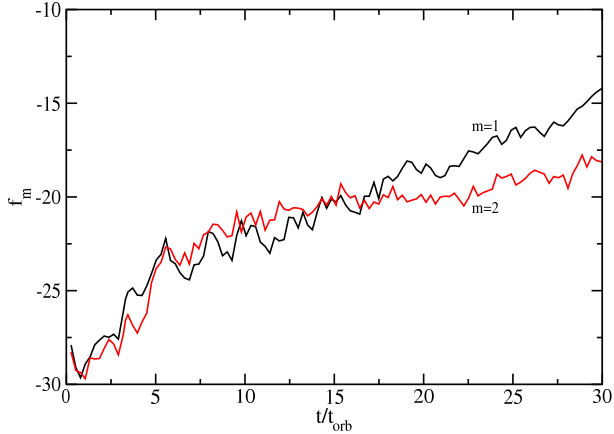}}
\caption{PPI mode growth.
\label{fig:PPI}}
\end{figure}

\begin{figure}
\scalebox{0.5}{\includegraphics{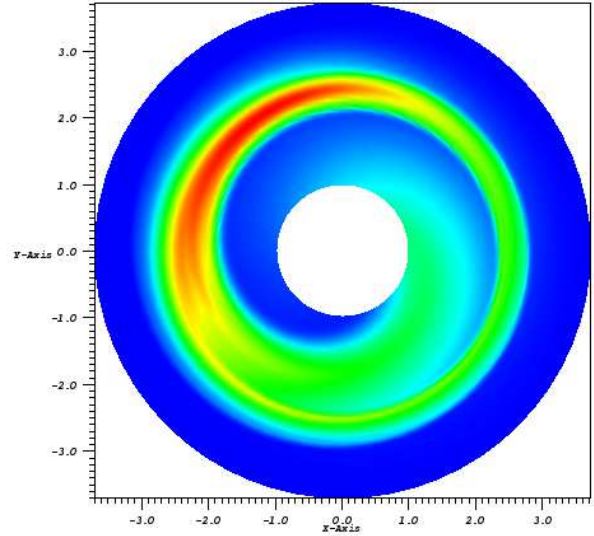}}
\caption{Density in the midplane of the torus.  This plot is shown 
using the logarithmic radial coordinate.
\label{fig:PPIden}}
\end{figure}

It is useful for this simulation to add a diagnostic 
to track the growth of the main PPI modes.  
We extract the $m=1$ and 2 modes by computing azimuthal density averages as 
\citep{dev02}
\begin{eqnarray}
\mathrm{Re}[k_m(r)] = \int_0^{2\pi} \rho (r,\pi/2,\phi)
            \cos(m\phi) d \phi ~, \\
\mathrm{Im}[k_m(r)] = \int_0^{2\pi} \rho (r,\pi/2,\phi)
            \sin(m\phi) d \phi ~.
\end{eqnarray}
The power in mode $m$ is then
\begin{eqnarray}
f_m & = & \frac{1}{r_{max}-r_{min}}\int_{r_{min}}^{r_{max}} \ln
    \left[\left(\left\{\mathrm{Re}[k_m(r)]\right\}^2 \right. \right. \\ \nonumber
  & & {}+ \left. \left. \left\{\mathrm{Im}[k_m(r)]\right\}^2\right)\right] d r ~,
\end{eqnarray}
where $r_{min}$ and $r_{max}$ are the approximate inner and outer
edges of the disk.  
In Figure \ref{fig:PPI} we show the growth of these 
two modes as a function of time.  As expected, there is significant 
mode growth for this slender torus.  Furthermore, this mode 
growth does not appear to saturate before the end of the simulation.
Figure \ref{fig:PPIden} shows the midplane density of this model 
at $t/t_{orb}=33$.  An $m=1$ ``planet'' and inflowing spiral wave are both 
apparent.

The question of interest now is what effect this PPI growth has on 
the {\it p}-mode oscillations.  From Figure 
\ref{fig:mdot} it is clear that the PPI growth coincides with a significant 
increase in the mass accretion rate, due to the angular momentum 
transport of the spiral waves generated in the torus.  Despite the very 
significant changes to the structure of the torus, the fundamental 
{\it p}-mode is able to survive at least to the end of this simulation, 
as is apparent in Figure \ref{fig:mdot}.  However, the power in this 
mode is significantly reduced and it is smeared out in frequency space 
as shown in Figure \ref{fig:powermdot}.  The frequency broadening is 
probably due to changes in the internal structure of the torus.  The 
damping is probably a result of leaking wave energy into the black hole 
along the accretion stream.

\subsection{2D Axisymmetric MHD \label{sec:2dmhd}}

I now explore the effect of adding initially weak poloidal magnetic field 
loops to the axisymmetric torus investigated in \S \ref{sec:2dhydro}.  
The presence of the poloidal field triggers the violent growth of the 
magneto-rotational instability \citep[MRI][]{bal91,haw91}.  
The MRI facilitates angular momentum transport through 
magnetohydrodynamic turbulence.  Similar to 
\S \ref{sec:3dhydro}, the focus here 
is to determine whether the {\it p}-mode oscillations seen 
in the 2D hydrodynamic 
simulation can survive in the presence of a known instability, in 
this case the MRI.

The initial magnetic field vector potential is \citep{dev03a}
\begin{equation}
A_\phi = \left\{ \begin{array}{ccc}
		  k(\rho-\rho_{min}) & \mathrm{for} & \rho\ge\rho_{min}~, \\
		  0                  & \mathrm{for} & \rho<\rho_{min}~.
	 	 \end{array} \right. 
\end{equation}
The non-zero spatial magnetic field components are then 
$\mathcal{B}^r = - \partial_\theta A_\phi$ and 
$\mathcal{B}^\theta = \partial_r A_\phi$.  These poloidal field loops 
coincide with the isodensity contours of the torus.  The constant $k$ 
is normalized such that initially 
$\beta=P/(\vert \vert B \vert \vert^2/8\pi) \ge 100$ 
throughout the torus.  
The parameter $\rho_{min}=0.5*\rho_{center}$ is used to keep the field 
a suitable distance inside the surface of the torus.

\begin{figure}
\scalebox{0.5}{\includegraphics{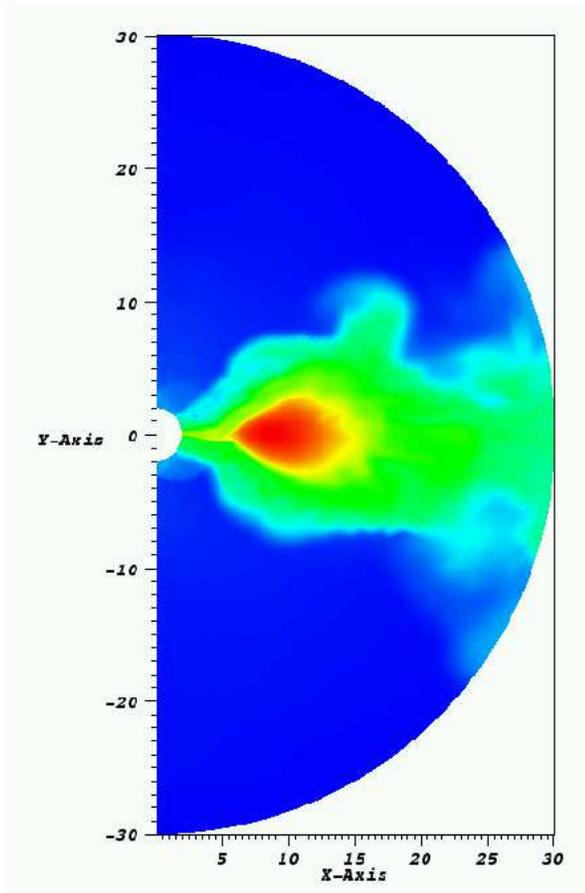}}
\caption{Logarithm of density for the initial and final torus states 
for the 2D MHD simulation.
\label{fig:2dm_den}}
\end{figure}

The presence of MRI turbulence dramatically increases the mass 
accretion rate compared to the hydrodynamic torus 
(see Figure \ref{fig:mdot}) 
and also redistributes the disk material into a broader, more radially 
extended structure as shown in Figure \ref{fig:2dm_den}.
It is important to note that axisymmetric MRI 
simulations are susceptible 
to a particularly violent form of the poloidal MRI, called the channel 
solution \citep{haw92}, which can itself lead to highly episodic 
mass accretion \citep[cf.][]{dev03b}, apparent in Figure 
\ref{fig:mdot}.

Comparisons of the power spectra of the $L_2$ density norms 
(Figure \ref{fig:powerrho}) and mass accretion rates 
(Figure \ref{fig:powermdot}) suggest that 
the violent nature of the channel solution 
strongly damps the 
{\it p}-mode oscillations.  The accretion spectrum of this simulation 
shows no power in the fundamental mode or its overtones.  
The $L_2$ density norm spectrum shows 
some power in the fundamental mode, although the amplitude of the 
oscillation decays with time, as 
apparent in Figure \ref{fig:l2_rho}.  It remains to be seen if this 
strong damping is restricted to the onset of the MRI or is a more 
general feature of these turbulent tori.



\section{Summary \label{sec:summary}}

An obvious extension of this work will be to perform 3D MHD simulations.  
There are two potentially important differences between magnetized 
axisymmetric (2D) and nonaxisymmetric (3D) tori.  One is the fact 
that the very violent channel solution of MRI (seen in Figure 
\ref{fig:2dm_den}) is itself susceptible 
to a nonaxisymmetric instability that destroys its coherence \citep{goo94}.  
As a consequence, accretion from a nonaxisymmetric torus is much less 
episodic than the equivalent 2D torus \citep{dev03b}.  
The second is that the MRI is 
unable to sustain itself through dynamo action in axisymmetry.  The 
sustained activity of the MRI in three-dimensions will be important in 
assessing the importance of {\it p}-mode oscillations in 
magnetized relativistic tori.

If acoustic modes are indeed strongly damped in turbulent MRI disks, then 
the puzzle of QPOs in black-hole X-ray binaries becomes more 
problematic.  The MRI only requires that a disk be differentially 
rotating at a rate that decreases with radial distance and that it have 
a weak magnetic field.  Both of these conditions 
are thought to be generally met in astrophysical accretion disks.  
In light of this and the results presented here, it appears that 
{\it p}-mode oscillations {\it may not} be a viable mechanism for 
generating observable QPOs in realistic accretion disks, but 3D MHD 
simulations will be necessary to confirm this.

{\it Animated movies of these results are available at \\
http://www.physics.ucsb.edu/$\sim$fragile/research.html.}

\begin{acknowledgments}
I would like to thank the VisIt development team at
Lawrence Livermore National Laboratory (http://www.llnl.gov/visit/),
in particular Hank Childs, for visualization support.
I would also like to acknowledge many useful discussions on this
work, particularly with L. Rezzolla, O. Zanotti, O. Blaes, and P. Anninos.
This work was partially performed
under the auspices of the U.S. Department of Energy by
University of California, Lawrence
Livermore National Laboratory under Contract W-7405-Eng-48.  
Funding support was also provided by NSF grant AST 0307657.
\end{acknowledgments}


\bibliographystyle{apsrev}
\bibliography{myrefs}

\begin{thebibliography}{15}
\expandafter\ifx\csname natexlab\endcsname\relax\def\natexlab#1{#1}\fi
\expandafter\ifx\csname bibnamefont\endcsname\relax
  \def\bibnamefont#1{#1}\fi
\expandafter\ifx\csname bibfnamefont\endcsname\relax
  \def\bibfnamefont#1{#1}\fi
\expandafter\ifx\csname citenamefont\endcsname\relax
  \def\citenamefont#1{#1}\fi
\expandafter\ifx\csname url\endcsname\relax
  \def\url#1{\texttt{#1}}\fi
\expandafter\ifx\csname urlprefix\endcsname\relax\def\urlprefix{URL }\fi
\providecommand{\bibinfo}[2]{#2}
\providecommand{\eprint}[2][]{\url{#2}}

\bibitem[{\citenamefont{Nowak and Wagoner}(1991)}]{now91}
\bibinfo{author}{\bibfnamefont{M.~A.} \bibnamefont{Nowak}} \bibnamefont{and}
  \bibinfo{author}{\bibfnamefont{R.~V.} \bibnamefont{Wagoner}},
  \bibinfo{journal}{\apj} \textbf{\bibinfo{volume}{378}}, \bibinfo{pages}{656}
  (\bibinfo{year}{1991}).

\bibitem[{\citenamefont{{Zanotti} et~al.}(2003)\citenamefont{{Zanotti},
  {Rezzolla}, and {Font}}}]{zan03}
\bibinfo{author}{\bibfnamefont{O.}~\bibnamefont{{Zanotti}}},
  \bibinfo{author}{\bibfnamefont{L.}~\bibnamefont{{Rezzolla}}},
  \bibnamefont{and} \bibinfo{author}{\bibfnamefont{J.~A.}
  \bibnamefont{{Font}}}, \bibinfo{journal}{\mnras}
  \textbf{\bibinfo{volume}{341}}, \bibinfo{pages}{832} (\bibinfo{year}{2003}).

\bibitem[{\citenamefont{{Rezzolla} et~al.}(2003)\citenamefont{{Rezzolla},
  {Yoshida}, {Maccarone}, and {Zanotti}}}]{rez03}
\bibinfo{author}{\bibfnamefont{L.}~\bibnamefont{{Rezzolla}}},
  \bibinfo{author}{\bibfnamefont{S.}~\bibnamefont{{Yoshida}}},
  \bibinfo{author}{\bibfnamefont{T.~J.} \bibnamefont{{Maccarone}}},
  \bibnamefont{and}
  \bibinfo{author}{\bibfnamefont{O.}~\bibnamefont{{Zanotti}}},
  \bibinfo{journal}{\mnras} \textbf{\bibinfo{volume}{344}},
  \bibinfo{pages}{L37} (\bibinfo{year}{2003}).

\bibitem[{\citenamefont{{Zanotti} et~al.}(2005)\citenamefont{{Zanotti}, {Font},
  {Rezzolla}, and {Montero}}}]{zan05}
\bibinfo{author}{\bibfnamefont{O.}~\bibnamefont{{Zanotti}}},
  \bibinfo{author}{\bibfnamefont{J.~A.} \bibnamefont{{Font}}},
  \bibinfo{author}{\bibfnamefont{L.}~\bibnamefont{{Rezzolla}}},
  \bibnamefont{and} \bibinfo{author}{\bibfnamefont{P.~J.}
  \bibnamefont{{Montero}}}, \bibinfo{journal}{\mnras}
  \textbf{\bibinfo{volume}{356}}, \bibinfo{pages}{1371} (\bibinfo{year}{2005}).

\bibitem[{\citenamefont{{De Villiers} and
  {Hawley}}(2003{\natexlab{a}})}]{dev03a}
\bibinfo{author}{\bibfnamefont{J.}~\bibnamefont{{De Villiers}}}
  \bibnamefont{and} \bibinfo{author}{\bibfnamefont{J.~F.}
  \bibnamefont{{Hawley}}}, \bibinfo{journal}{\apj}
  \textbf{\bibinfo{volume}{589}}, \bibinfo{pages}{458}
  (\bibinfo{year}{2003}{\natexlab{a}}).

\bibitem[{\citenamefont{{Papadopoulos} and {Font}}(1998)}]{pap98}
\bibinfo{author}{\bibfnamefont{P.}~\bibnamefont{{Papadopoulos}}}
  \bibnamefont{and} \bibinfo{author}{\bibfnamefont{J.~A.}
  \bibnamefont{{Font}}}, \bibinfo{journal}{\prd} \textbf{\bibinfo{volume}{58}},
  \bibinfo{pages}{24005} (\bibinfo{year}{1998}).

\bibitem[{\citenamefont{{Font} et~al.}(1998)\citenamefont{{Font}, {Ib{\' a}{\~
  n}ez}, and {Papadopoulos}}}]{fon98b}
\bibinfo{author}{\bibfnamefont{J.~A.} \bibnamefont{{Font}}},
  \bibinfo{author}{\bibfnamefont{J.~M.~.} \bibnamefont{{Ib{\' a}{\~ n}ez}}},
  \bibnamefont{and}
  \bibinfo{author}{\bibfnamefont{P.}~\bibnamefont{{Papadopoulos}}},
  \bibinfo{journal}{\apjl} \textbf{\bibinfo{volume}{507}}, \bibinfo{pages}{L67}
  (\bibinfo{year}{1998}).

\bibitem[{\citenamefont{{Michel}}(1972)}]{mic72}
\bibinfo{author}{\bibfnamefont{F.~C.} \bibnamefont{{Michel}}},
  \bibinfo{journal}{\apss} \textbf{\bibinfo{volume}{15}}, \bibinfo{pages}{153}
  (\bibinfo{year}{1972}).

\bibitem[{\citenamefont{{Kato}}(2001)}]{kat01}
\bibinfo{author}{\bibfnamefont{S.}~\bibnamefont{{Kato}}},
  \bibinfo{journal}{\pasj} \textbf{\bibinfo{volume}{53}}, \bibinfo{pages}{1}
  (\bibinfo{year}{2001}).

\bibitem[{\citenamefont{{De Villiers} and {Hawley}}(2002)}]{dev02}
\bibinfo{author}{\bibfnamefont{J.}~\bibnamefont{{De Villiers}}}
  \bibnamefont{and} \bibinfo{author}{\bibfnamefont{J.~F.}
  \bibnamefont{{Hawley}}}, \bibinfo{journal}{\apj}
  \textbf{\bibinfo{volume}{577}}, \bibinfo{pages}{866} (\bibinfo{year}{2002}).

\bibitem[{\citenamefont{{Balbus} and {Hawley}}(1991)}]{bal91}
\bibinfo{author}{\bibfnamefont{S.~A.} \bibnamefont{{Balbus}}} \bibnamefont{and}
  \bibinfo{author}{\bibfnamefont{J.~F.} \bibnamefont{{Hawley}}},
  \bibinfo{journal}{\apj} \textbf{\bibinfo{volume}{376}}, \bibinfo{pages}{214}
  (\bibinfo{year}{1991}).

\bibitem[{\citenamefont{{Hawley}}(1991)}]{haw91}
\bibinfo{author}{\bibfnamefont{J.~F.} \bibnamefont{{Hawley}}},
  \bibinfo{journal}{\apj} \textbf{\bibinfo{volume}{381}}, \bibinfo{pages}{496}
  (\bibinfo{year}{1991}).

\bibitem[{\citenamefont{{Hawley} and {Balbus}}(1992)}]{haw92}
\bibinfo{author}{\bibfnamefont{J.~F.} \bibnamefont{{Hawley}}} \bibnamefont{and}
  \bibinfo{author}{\bibfnamefont{S.~A.} \bibnamefont{{Balbus}}},
  \bibinfo{journal}{\apj} \textbf{\bibinfo{volume}{400}}, \bibinfo{pages}{595}
  (\bibinfo{year}{1992}).

\bibitem[{\citenamefont{{De Villiers} and
  {Hawley}}(2003{\natexlab{b}})}]{dev03b}
\bibinfo{author}{\bibfnamefont{J.}~\bibnamefont{{De Villiers}}}
  \bibnamefont{and} \bibinfo{author}{\bibfnamefont{J.~F.}
  \bibnamefont{{Hawley}}}, \bibinfo{journal}{\apj}
  \textbf{\bibinfo{volume}{592}}, \bibinfo{pages}{1060}
  (\bibinfo{year}{2003}{\natexlab{b}}).

\bibitem[{\citenamefont{{Goodman} and {Xu}}(1994)}]{goo94}
\bibinfo{author}{\bibfnamefont{J.}~\bibnamefont{{Goodman}}} \bibnamefont{and}
  \bibinfo{author}{\bibfnamefont{G.}~\bibnamefont{{Xu}}},
  \bibinfo{journal}{\apj} \textbf{\bibinfo{volume}{432}}, \bibinfo{pages}{213}
  (\bibinfo{year}{1994}).

\end{thebibliography}

\end{document}